


\magnification=\magstep1
\baselineskip 20pt



\input jnl.tex
\input reforder.tex
\input eqnorder.tex



\twelvepoint
\oneandahalfspace
\doublespace
\firstpageno=2
\tolerance=2000

\title Spin texture in weakly doped $CuO_2$ planes
explaining $\xi (x,T)$ and Raman scattering experiments
\bigskip
\bigskip

\author R.J. Gooding
\affil Department of Physics, Queen's University,
Kingston, Ontario, Canada K7L 3N6
\bigskip
\bigskip
\author A. Mailhot
\affil Centre de Recherche en Physique du Solide et D\'epartement de Physique,
Universit\'e de Sherbrooke,
Sherbrooke, Qu\'ebec, Canada J1K 2R1

\abstract
A model of $CuO_2$ planes weakly doped with partially
delocalised holes is considered. The effect of such a hole on the background
AFM
spin texture can be represented by a purely magnetic Hamiltonian
$H = - \sum_{(ijk)} (\vec S_i \cdot \vec S_j \times \vec S_k)^2$,
where the summation is over the four triangles of a single plaquette.
We show that this model of randomly distributed chiral spin defects leads to
an in--plane spin correlation length
approximately described by $\xi^{-1} (x,T) = \xi^{-1} (0,T) + \xi^{-1} (x,0)$,
consistent with neutron scattering experiments on $La_{2-x}Sr_xCuO_4$.
Further, this model leads to favourable comparisons with
$B_{1g}$ Raman scattering results for the
same cuprate system.

\endtopmatter

\beginparmode
\twelvepoint
\doublespace

Many detailed experiments have recently been conducted examining
the magnetic properties of the weakly doped Bedn\"orz--M\"uller
high $T_c$ compound, $La_{2-x}(Ba,Sr)_xCuO_4$. Consequently,
models purportedly representing the charge and spin densities
in these systems can be scrutinized by comparing their predictions
with experiment.  In this letter we
shall compare the predictive powers of
our model to the following 2 studies: (i) 2--axis
elastic neutron scattering studies\refto{Keimer} of the in--plane
correlation length as a
function of doping and temperature, viz. $\xi (x,T)$. These authors
found that the effects of doping and temperature combined in a unusual
fashion, such that an empirical fit to the correlation length
was achieved via $\xi^{-1} (x,T) = \xi^{-1} (x,0) + \xi^{-1} (0,T)$.
(ii) Raman scattering\refto{Sugai} in a geometry
appropriate for studying the $B_{1g}$ channel. These authors found
that the position of the broad triangular peak associated with the two--magnon
response was slowly shifted to lower energies with increasing
doping.

In order to understand
the magnetic background that is responsible for the above results,
we first consider the perturbation that doping
produces in the $CuO_2$ planes, and then model it in a simple fashion.
All of the above work focused on the
low--doping regime of these materials, viz.  $0 {< \atop \sim}
x {< \atop \sim} .05$. For this range of $x$, both $\mu^+ sR$
relaxation\refto{Harshman} and
resistivity measurements\refto{Hayden} indicate that the carriers are at least
partially localised. To be specific, work by Harshman \etal\refto{Harshman}
showed that in  samples with $x = .01, .02,$ and even $.05$, spin
freezing occurs below some temperature, with the field inhomogeneity
substantially growing with $x$. Thus, a considerably inhomogeneous
distribution of perturbing $O^-$ holes may be inferred. This conclusion
is also consistent with the insulating
properties of these systems at low temperatures.\refto{Hayden}

Then, one may argue that the holes will not be completely localised on
a single O site by appealing to an argument made (in a different context) by
Halperin and Varma,\refto{centralpeak} viz. that a defect in a
sample will fluctuate in an attempt to restore the
symmetry of the pure system.  Further, one may quantify the degree of
localisation by noting that a single O hole in a $CuO_2$ plane is
produced by the replacement of a trivalent $La$ ion with a divalent
$Sr$ - the hole feels a pinning potential, and as discussed
by Rabe and Bhatt,\refto{BhattRabe} in such a situation strong localization of
the charge density (perhaps one or two lattice constants) is found.

The simplest model of a hole that restores the square symmetry of
the lattice, and yet is still largely localised,
is one in which the hole is free to move around a single
plaquette, and it is this representation of the weakly
doped state that we shall study. One of us has previously examined
this problem,\refto{skyrmion} and found that the ground state of this
situation corresponds to a novel state in which the hole circulates
with a nonzero charge current density, in either a clockwise or
counter--clockwise direction around the plaquette.
Then, as a consequence of the
coupling of the spin and charge current densities,\refto{ShraimanSiggia}
the AFM background is
perturbed in a fashion reminiscent of an excitation of the 2D classical
Heisenberg model known as a skyrmion\refto{Belavin} with a topological
charge of $\pm 1$.
The perturbation on the magnetic state was found to be
long--ranged,\refto{skyrmion} which is also consistent with the cluster studies
of Bhatt and Rabe,\refto{BhattRabe} who found that while charge is localised
by the impurity potential, the spin deviation is not. A discussion
detailing the interaction responsible for this (locally) chiral spin
texture may be found in Ref. 7.

We now consider the inclusion of a non--zero density of such partially
localised holes into the 2D lattice. While the single hole problem
may be treated within the familiar context of a $t-J$ model, this is
not feasible for N carriers - each localised in a given region of
the plane chosen, these regions being distributed in the
plane at random. Instead, we shall appeal to two ideas to produce
a useful effective Hamiltonian that represents this
system.

Firstly, it has been shown that the extreme quantum limit
($S = 1/2$) of the \2d Heisenberg model on a square lattice, viz.
$$
H_J =  J \sum_{(ij)} ~\vec S_i \cdot \vec S_j  \tag heisenberg
$$
where $(ij)$ denotes near neighbours, has
a correlation length described by the {\it classical} nonlinear sigma model,
which then has its bare spin stiffness $\rho^0_s$ and perpendicular
susceptibility $\chi^0_{\perp}$ dressed with quantum
fluctuations. In particular, this has
been demonstrated by extensive Monte Carlo studies
of Makivic and Ding.\refto{Makivic} Further, when such a system is doped,
the semiclassical theory of Shraiman and Siggia,\refto{ShraimanSiggia}
which is based on treating the spin degrees of freedom
classically, has had considerable success; e.g.,
their prediction of the spiral phase. The stability
of the ground state with a skyrmion--like spin texture
discussed above was also explained using a similar (in this
case, real space) semiclassical theory. Here we shall also treat
the spin degrees of freedom using classical spins, and ignore
the renormalizing affects introduced by quantum fluctuations.

Secondly, we shall eliminate the motional degree of freedom
of the hole, and replace it by a purely magnetic
interaction which produces the same disturbance of the
background N\'eel state that the hole produces.
This perturbation will be localised to only act on the
spins of a single plaquette. We denote the perturbing
interaction by $H_{int} (p)$, where $p$ represents the
spatial location of one plaquette.

Since we require that the ground state induced by this interaction
mimics the (locally) chiral ground state of the
single-hole-moving-around-a-single-plaquette problem,\refto{skyrmion}
it is natural that we focus on an interaction that leads to
a doubly degenerate chiral ground state.\refto{WWZ}
Localising this interaction to a single plaquette leads to
$$
\eqalign{
H_{int} (p) = - (D/S^4) \Big[
&(\vec S_i \cdot \vec S_j \times \vec S_k)^2 ~+~
(\vec S_j \cdot \vec S_k \times \vec S_{\ell})^2~+~\cr
&(\vec S_k \cdot \vec S_{\ell} \times \vec S_i)^2~+~
(\vec S_{\ell} \cdot \vec S_i \times \vec S_j)^2 \Big]\cr} \tag chiral
$$
where $i,j,k,$ and $\ell$ are the four corners of the $pth$ plaquette,
and $D$ is a strictly positive constant that we specify below (note
that our classical spins are chosen to be of length $S$). Then,
for a non--zero density of holes we are lead to consider
the following effective Hamiltonian:
$$
H_{eff} = H_J + {\sum_p}^\prime H_{int} (p) \tag effective
$$
where the primed summation implies that the average density of
holes is fixed, with the location of the perturbed plaquettes
chosen at random.

For classical vector spins it is trivial to show that for just one
plaquette perturbed by Eq.~\(chiral) the ground state is a doubly
degenerate state with a non--coplanar spin texture
reminiscent of the skyrmion problem (see Eq. (8) in Ref. 7).
Further, one may show, by evaluating the energy of this
ground state relative to the unperturbed N\'eel state as a function
of system size, that for $D > 1.2867$ this doubly degenerate
ground state is stable. (One may also show that for $D > 1.7493$
the first excited state, a non--coplanar zero chirality spin texture,
is also stable with respect to the N\'eel state - since an analogous
state (viz., a state with the same symmetry) was found to
be close in energy to the skyrmion ground state in the
single hole problem,\refto{skyrmion} we feel it is desirable to
include it here.) Lastly, examining how the AFM order parameter changes
due to the introduction of a single perturbed plaquette, again
with respect to system size, we find that the change grows
logarithmically with the size of the lattice. Similar arguments
were used previously by us\refto{ourprb} to demonstrate that this
implies that the perturbation of the AFM background is long--ranged.
Clearly, the classical spin texture induced by Eq.~\(chiral)
is qualitatively equivalent to that found for the single hole problem.

We have performed standard Metropolis Monte Carlo simulations
of $L \times L$ lattices, with L= 12, 18, 24, 30, and 36, for
{\it average} densities of holes corresponding to
$x = .02, .03$, and $.04$, using
Eq.~\(effective).  Beginning at infinite T, and
using a cooling rate of $.0005 J$ (equilibrating for
25,000 MCS at each temperature) the system
was cooled until temperatures around $J/2$ were reached. At this
lower $T$, the correlation length was found to be just greater than half the
system size. For each density of holes, an average over disorder
was affected by simulating 15 different distributions of
perturbed plaquettes; a few systems were averaged over 30 distributions,
but essentially no quantitative changes in our results for the spin--spin
correlation function were found. The correlation length was
extracted by fitting $C_r \equiv (-1)^r <\vec S_0 \cdot \vec S_r>$ to
$C_r + C_{L-r}$ to take into account finite size affects;\refto{Makivic}
the form
$$
C_r = A r^{-\nu} \exp{ -r/{\xi(x,T)}} \tag cofr
$$
was assumed to adequately define the long--distance correlation
function, and thus the doping and temperature dependent
in--plane spin correlation lengths $\xi(x,T)$ were extracted.
When performing this fit, only data for $C_r$ with $ 2, 3$ or
$4 < r < L/2 $ that led to
fits with a $\chi^2$ of less than $10^{-4}$ were considered to
be sufficiently well sampled to be accurate.
We are convinced that our simulations were approximately ergodic
for the undoped system - our low $T$ data exactly approached
the form expected from the low $T$ renormalization
theory,\refto{Polyakov} viz.
$$
\xi(0,T \approx 0) = C_0 {\exp {2\pi / T}}
\Big[ {1 \over 1 + 2\pi /T}\Big] \tag xi
$$
(note that $\rho^0_s = 1$ for the classical system),
while results very close to the high--temperature series expansion
prediction were found at higher T. Similar conclusions
were found in earlier MC work.\refto{Shenker} We found that for the doped
systems it was, of course, more difficult to obtain converged averages
for the spin--spin correlation function, and thus the very slow cooling
rate and long equilibration times were essential.

Some of our results for the undoped lattices are shown in Fig. 1;
a more detailed account will be presented elsewhere.
Empirical fits to $\xi(x,T)$ found in 2--axis elastic
neutron scattering data obtained by Keimer\refto{Keimer} led
to
$$
\eqalign{
\xi^{-1} (x,T) &= \xi^{-1} (x,0) + \xi^{-1} (0,T)\cr
\xi (x,0) &= a / \sqrt x\cr}\tag keimer
$$
where $a$ is the lattice constant. This result
is similar to our computer generated data.
However, we find that, more generally, $\xi (x,0) \sim x^{-f}$
with $ .5 \leq f \leq 1.$ leads to good fits of Eq. (6a)
to our data. For example, the data shown in Fig. 1 is
extremely well fit by $f = .75 \pm .15$.

In addition to the MC work, we have examined the
$T=0$ spin--spin correlation function; this supplements the above MC work
by providing one low T data point, and thus better tests how well
our simulations agree with Eq.~\(keimer).  We used a
technique identical to our earlier work\refto{ourprb} on an Aharony--type
of spin defect perturbing Eq.~\(heisenberg). Our results for
$\xi(x,T=0)$ at $x = .02, .03, .04$, and $.05$, are $20 \pm 8, 12 \pm 5,
9 \pm 4$, and $6 \pm 3$.  This data, within our error bars,
agrees with the form $\xi(x,0) \sim x^{-.65 \pm .1}$.

In the above work the value of the local chiral interaction $D$
does not seem to be important in reproducing the data of Keimer. Recall
from above that we required $D > 1.75$; thus,
we have studied $D = 2,3,4,$ and 5, and our $D = 3$ results
are those shown in Fig. 1. It is to be stressed that
$D = 2$ produced similar results.  Unfortunately, because
of metastability problems we were not able to obtain converged averages
for systems with $D > 3$, even
when much slower cooling rates were used. Therefore,
another experimentally inferred value of $D$ is discussed below.

The conclusion from the above results are (i)
our model of partially localised holes producing an
inhomogeneous non--coplanar
spin texture leads to a correlation length that agrees
with experiment, and (ii) weak doping
introduces a length scale approximately determined
by the average spacing between
randomly distributed holes, and then
each such region (possibly a domain) behaves like a gapless
Heisenberg model with an upper bound for the correlation length.

We now turn our attention to a second experiment\refto{Sugai} on weakly doped
$La_{2-x}(Ba,Sr)_xCuO_4$, viz. a study of the two--magnon scattering
for the $B_{1g}$ geometry. These authors found
that the broad triangular peak, which to lowest order may be
associated with the spin flip of a pair of nearest--neighbour spins,
is shifted to lower energies with increasing doping. At $T = 30 K$,
the peak for $x = 0.034$ is shifted down by 5 \% relative
to the $x = 0$ response.

In comparison to the undoped sample, the triangular peaks are somewhat
broadened.  When considered along with the $\mu^+$sR relaxation
result\refto{Harshman} for the field inhomogeneity of
the weakly doped systems, it may be inferred that the spin texture
may be due to an inhomogeneous spin arrangement, such as the one we
are proposing in this paper. We will not discuss the broadening
further. Instead, in order to assess the viability of Eq.~\(effective)
in describing the spin texture via a comparison to
the Raman scattering, we consider the spin state arising
in a mean--field treatment of this Hamiltonian, viz.
$$
H^{MFT}_{eff} = H_J + x \sum_p H_{int} (p) \tag mft
$$
where the summation is over {\it all} plaquettes. The use
of a mean--field type approximation is partly justified by
the long--ranged spin distortions imposed by Eq.~\(chiral).
Now, using the operator identity\refto{WWZ} for $S = {1\over 2}$ spins
$$
(\vec S_1 \cdot \vec S_2 \times \vec S_3)^2 =
{1\over 64} \Big[15 - 4 (\vec S_1 + \vec S_2 + \vec S_3)^2\Big]
\tag wwz
$$
$H^{MFT}_{eff}$ is seen to reduce to the
so--called $J - J^\prime$ model\refto{ChandraDoucot} representing
a frustrated AFM on a square lattice, where
$$
{J^\prime \over J }\sim {{D~x / (4~S^4)} \over
{1~+~D~x/(2~S^4)}}. \tag definingD
$$

We now use Eq.~\(definingD) to predict a value of $D$, the strength
of the chiral interaction appearing in Eq.~\(chiral), based on
a comparison to the Raman scattering.  Firstly, following Anderson's
spin--wave theory\refto{Anderson} for an AFM we replace $S^2 \rightarrow
S(S+1)$.
Secondly, Suzuki and Natsume\refto{Suzuki} have
used  a modified spin--wave theory for Raman scattering applied
to the $J - J^\prime$ model; using their results it follows that for
$J^\prime / J \sim .04 \pm .01 $ the 5 \% shift of the $x = .034$ peak position
found by Sugai \etal may be explained.\refto{Sugai}
Finally, Eq.~\(definingD) leads us to infer $D \sim 2.9 \pm .7$.
This is to be compared with the values of $D = 2$ and $3$ used in the
Monte Carlo work studying the spin--spin correlation function discussed
above.  Thus, even accounting for the crudeness of
our mean--field approximation leading
to Eq.~\(definingD), it seems that essentially {\it quantitative} agreement
with {\it both} the neutron and Raman scattering experiments is achieved
through the use of Eq.~\(effective).

Summarizing, we have considered the weakly doped
$CuO_2$ planes in which the $O^-$ holes are modeled by partially
delocalised holes that produce a locally chiral spin texture
reminiscent of a singly charged skyrmion.\refto{skyrmion} The motional
degree of freedom of the holes are then replaced by a purely
magnetic interaction whose ground state (and first excited state)
mimics the problem that includes the hole, and a non--zero
density of such defects are easily incorporated into a square lattice.
Then, appealing to
the success of the classical Heisenberg model in reproducing the
spin correlation length of its extreme quantum
counterpart, we have performed classical Monte Carlo simulations
which seemingly reproduce the empirical relation found by
Keimer \etal\refto{Keimer} stated in Eq.~\(keimer); the question
of how quantum fluctuations affect the disordered N\'eel state
has not been considered.  Lastly, the strength
of the local chiral interaction $D$ in Eq.~\(chiral) that reproduces the
behaviour of $\xi(x,T)$ is the same as that required for
a mean--field theory of the same Hamiltonian when compared to
the peak shifts found in $B_{1g}$ Raman scattering.

It will be interesting to see how well this model reproduces other
experimental results, such as
static susceptibility\refto{Johnston} measurements, which also probe
the magnetic spin texture of the weakly doped $CuO_2$ planes.
Work is presently in progress addressing this issue.

\bigskip
We wish to thank Ferdinand Borsa, Stuart Trugman, and especially
Dave Johnston,
for valuable comments.  This work was supported by the NSERC of Canada, and
the Fonds pour la Formation de Chercheurs et l'Aide \`a la
Recherche du Qu\'ebec.
\endpage
\centerline{Figure Captions:}

\item{1.} The inverse correlation length $\xi^{-1} (x,T)$, relative
to the lattice constant $a$, plotted
as a function of temperature T
for $x = 0, .02$, and $.04$; the simulations were conducted
on a $30 \times 30$ lattice. The constant $D$ in Eq.~\(chiral)
was chosen to be 3. A few representative error bars are shown.
\endpage

\references

\parskip=5pt plus 2pt

\refis{Keimer} B. Keimer, \etal, \prl 67, 1930, 1991;
Physica B {\bf 180,181} 15 (1992).

\refis{Sugai} S. Sugai, \etal, \prb 38, 6436, 1988.

\refis{Harshman} D.R. Harshman, \etal, \prb 38, 852, 1988.

\refis{Hayden} S.M. Hayden, \etal, \prl 66, 821, 1991.

\refis{centralpeak} B.I. Halperin, and C.M. Varma, \prb 14, 4030, 1976.

\refis{BhattRabe} K.M. Rabe, and R.N. Bhatt, \jap, 69, 4508, 1991.

\refis{skyrmion} R.J. Gooding, \prl 66, 2266, 1991.

\refis{ShraimanSiggia} B.I. Shraiman, and E.D. Siggia, \prl 61, 467, 1988;
\prl 62, 1564, 1989.

\refis{Belavin} A.A. Belavin, and P.M. Polyakov, JETP Lett.{\bf 22}, 245
(1975).

\refis{Makivic} M. Makivic, and H. Ding, \prb 43, 3563, 1991.

\refis{WWZ} X.G. Wen, \etal, \prb 39, 11413, 1989.

\refis{ourprb} R.J. Gooding, and A. Mailhot, \prb 44, 11852, 1991.

\refis{Polyakov} P.M. Polyakov, Phys. Lett. B {\bf 59}, 79 (1975).

\refis{Shenker} S.H. Shenker, and J. Tobochnik, \prb 22, 4462, 1980.

\refis{ChandraDoucot} P. Chandra, and B. Doucot, \prb 38, 6631, 1988.

\refis{Anderson} P.W. Anderson, \pr 86, 694, 1952.

\refis{Suzuki} T. Suzuki, and Y. Natsume, J. Phys. Soc. Jpn. {\bf 61}, 998
(1992).

\refis{Johnston} J.H. Cho, \etal, (submitted to Phys. Rev. Lett).

\endreferences

\endpage

\catcode`@=11
\newcount\r@fcount \r@fcount=0
\newcount\r@fcurr
\immediate\newwrite\reffile
\newif\ifr@ffile\r@ffilefalse
\def\w@rnwrite#1{\ifr@ffile\immediate\write\reffile{#1}\fi\message{#1}}

\def\writer@f#1>>{}
\def\referencefile{
  \r@ffiletrue\immediate\openout\reffile=\jobname.ref%
  \def\writer@f##1>>{\ifr@ffile\immediate\write\reffile%
    {\noexpand\refis{##1} = \csname r@fnum##1\endcsname = %
     \expandafter\expandafter\expandafter\strip@t\expandafter%
     \meaning\csname r@ftext\csname r@fnum##1\endcsname\endcsname}\fi}%
  \def\strip@t##1>>{}}

\def\citeall#1{\xdef#1##1{#1{\noexpand\cite{##1}}}}
\def\cite#1{\each@rg\citer@nge{#1}}	

\def\each@rg#1#2{{\let\thecsname=#1\expandafter\first@rg#2,\end,}}
\def\first@rg#1,{\thecsname{#1}\apply@rg}	
\def\apply@rg#1,{\ifx\end#1\let\next=\relax
\else,\thecsname{#1}\let\next=\apply@rg\fi\next}

\def\citer@nge#1{\citedor@nge#1-\end-}	
\def\citer@ngeat#1\end-{#1}
\def\citedor@nge#1-#2-{\ifx\end#2\r@featspace#1 
  \else\citel@@p{#1}{#2}\citer@ngeat\fi}	
\def\citel@@p#1#2{\ifnum#1>#2{\errmessage{Reference range #1-#2\space is bad.}
    \errhelp{If you cite a series of references by the notation M-N, then M and
    N must be integers, and N must be greater than or equal to M.}}\else%
 {\count0=#1\count1=#2\advance\count1
by1\relax\expandafter\r@fcite\the\count0,%
  \loop\advance\count0 by1\relax
    \ifnum\count0<\count1,\expandafter\r@fcite\the\count0,%
  \repeat}\fi}

\def\r@featspace#1#2 {\r@fcite#1#2,}	
\def\r@fcite#1,{\ifuncit@d{#1}		
    \expandafter\gdef\csname r@ftext\number\r@fcount\endcsname%
    {\message{Reference #1 to be supplied.}\writer@f#1>>#1 to be supplied.\par
     }\fi%
  \csname r@fnum#1\endcsname}

\def\ifuncit@d#1{\expandafter\ifx\csname r@fnum#1\endcsname\relax%
\global\advance\r@fcount by1%
\expandafter\xdef\csname r@fnum#1\endcsname{\number\r@fcount}}

\let\r@fis=\refis			
\def\refis#1#2#3\par{\ifuncit@d{#1}
    \w@rnwrite{Reference #1=\number\r@fcount\space is not cited up to now.}\fi%
  \expandafter\gdef\csname r@ftext\csname r@fnum#1\endcsname\endcsname%
  {\writer@f#1>>#2#3\par}}

\def\r@ferr{\endreferences\errmessage{I was expecting to see
\noexpand\endreferences before now;  I have inserted it here.}}
\let\r@ferences=\references
\def\references{\r@ferences\def\endmode{\r@ferr\par\endgroup}}

\let\endr@ferences=\endreferences
\def\endreferences{\r@fcurr=0
  {\loop\ifnum\r@fcurr<\r@fcount
    \advance\r@fcurr by 1\relax\expandafter\r@fis\expandafter{\number\r@fcurr}%
    \csname r@ftext\number\r@fcurr\endcsname%
  \repeat}\gdef\r@ferr{}\endr@ferences}


\let\r@fend=\endpaper\gdef\endpaper{\ifr@ffile
\immediate\write16{Cross References written on []\jobname.REF.}\fi\r@fend}

\catcode`@=12

\citeall\refto		
\citeall\ref		%
\citeall\Ref		%

\catcode`@=11
\newcount\tagnumber\tagnumber=0

\immediate\newwrite\eqnfile
\newif\if@qnfile\@qnfilefalse
\def\write@qn#1{}
\def\writenew@qn#1{}
\def\w@rnwrite#1{\write@qn{#1}\message{#1}}
\def\@rrwrite#1{\write@qn{#1}\errmessage{#1}}

\def\taghead#1{\gdef\t@ghead{#1}\global\tagnumber=0}
\def\t@ghead{}

\expandafter\def\csname @qnnum-3\endcsname
  {{\t@ghead\advance\tagnumber by -3\relax\number\tagnumber}}
\expandafter\def\csname @qnnum-2\endcsname
  {{\t@ghead\advance\tagnumber by -2\relax\number\tagnumber}}
\expandafter\def\csname @qnnum-1\endcsname
  {{\t@ghead\advance\tagnumber by -1\relax\number\tagnumber}}
\expandafter\def\csname @qnnum0\endcsname
  {\t@ghead\number\tagnumber}
\expandafter\def\csname @qnnum+1\endcsname
  {{\t@ghead\advance\tagnumber by 1\relax\number\tagnumber}}
\expandafter\def\csname @qnnum+2\endcsname
  {{\t@ghead\advance\tagnumber by 2\relax\number\tagnumber}}
\expandafter\def\csname @qnnum+3\endcsname
  {{\t@ghead\advance\tagnumber by 3\relax\number\tagnumber}}

\def\equationfile{%
  \@qnfiletrue\immediate\openout\eqnfile=\jobname.eqn%
  \def\write@qn##1{\if@qnfile\immediate\write\eqnfile{##1}\fi}
  \def\writenew@qn##1{\if@qnfile\immediate\write\eqnfile
    {\noexpand\tag{##1} = (\t@ghead\number\tagnumber)}\fi}
}

\def\callall#1{\xdef#1##1{#1{\noexpand\call{##1}}}}
\def\call#1{\each@rg\callr@nge{#1}}

\def\each@rg#1#2{{\let\thecsname=#1\expandafter\first@rg#2,\end,}}
\def\first@rg#1,{\thecsname{#1}\apply@rg}
\def\apply@rg#1,{\ifx\end#1\let\next=\relax%
\else,\thecsname{#1}\let\next=\apply@rg\fi\next}

\def\callr@nge#1{\calldor@nge#1-\end-}
\def\callr@ngeat#1\end-{#1}
\def\calldor@nge#1-#2-{\ifx\end#2\@qneatspace#1 %
  \else\calll@@p{#1}{#2}\callr@ngeat\fi}
\def\calll@@p#1#2{\ifnum#1>#2{\@rrwrite{Equation range #1-#2\space is bad.}
\errhelp{If you call a series of equations by the notation M-N, then M and
N must be integers, and N must be greater than or equal to M.}}\else%
 {\count0=#1\count1=#2\advance\count1
by1\relax\expandafter\@qncall\the\count0,%
  \loop\advance\count0 by1\relax%
    \ifnum\count0<\count1,\expandafter\@qncall\the\count0,%
  \repeat}\fi}

\def\@qneatspace#1#2 {\@qncall#1#2,}
\def\@qncall#1,{\ifunc@lled{#1}{\def\next{#1}\ifx\next\empty\else
  \w@rnwrite{Equation number \noexpand\(>>#1<<) has not been defined yet.}
  >>#1<<\fi}\else\csname @qnnum#1\endcsname\fi}

\let\eqnono=\eqno
\def\eqno(#1){\tag#1}
\def\tag#1$${\eqnono(\displayt@g#1 )$$}

\def\aligntag#1\endaligntag
  $${\gdef\tag##1\\{&(##1 )\cr}\eqalignno{#1\\}$$
  \gdef\tag##1$${\eqnono(\displayt@g##1 )$$}}

\def\eqalignno#1{\displ@y \tabskip\centering
  \halign to\displaywidth{\hfil$\displaystyle{##}$\tabskip\z@skip
    &$\displaystyle{{}##}$\hfil\tabskip\centering
    &\llap{$\displayt@gpar##$}\tabskip\z@skip\crcr
    #1\crcr}}

\def\displayt@gpar(#1){(\displayt@g#1 )}

\def\displayt@g#1 {\rm\ifunc@lled{#1}\global\advance\tagnumber by1
        {\def\next{#1}\ifx\next\empty\else\expandafter
        \xdef\csname @qnnum#1\endcsname{\t@ghead\number\tagnumber}\fi}%
  \writenew@qn{#1}\t@ghead\number\tagnumber\else
        {\edef\next{\t@ghead\number\tagnumber}%
        \expandafter\ifx\csname @qnnum#1\endcsname\next\else
        \w@rnwrite{Equation \noexpand\tag{#1} is a duplicate number.}\fi}%
  \csname @qnnum#1\endcsname\fi}

\def\ifunc@lled#1{\expandafter\ifx\csname @qnnum#1\endcsname\relax}

\let\@qnend=\end\gdef\end{\if@qnfile
\immediate\write16{Equation numbers written on []\jobname.EQN.}\fi\@qnend}

\catcode`@=12
